\documentclass[twocolumn,preprintnumbers,amsmath,amssymb]{revtex4}
\usepackage{latexsym,amssymb,amsthm,amsmath,epsfig}

\begin{document}

\title{Two-body relaxation in a Fermi gas at a $p$-wave Feshbach resonance}
\author{Muhammad Waseem$^{1,2,3}$}
\email{waseem@ils.uec.ac.jp}
\author{Taketo Saito$^{1,2}$}
\author{Jun Yoshida$^{1,2}$}
\author{Takashi Mukaiyama$^{3}$}
\affiliation{%
$^{1}$\mbox{Department of Engineering Science, University of Electro-Communications, Tokyo 182-8585, Japan}\\
$^{2}$\mbox{Institute for Laser Science,University of Electro-Communications, Chofugaoka, Chofu, Tokyo 182-8585, Japan}\\
$^{3}$\mbox{Graduate School of Engineering Science, Osaka University, Machikaneyama, Toyonaka, Osaka 560-8531 Japan}\\
}
\date{\today }

\begin{abstract}
We systematically studied the two-body loss in a two-component Fermi gas of $^6$Li atoms near a $p$-wave Feshbach resonance. The two-body loss rate constants were measured for various temperatures and magnetic fields using atoms trapped in three-dimensional and quasi-two-dimensional traps. Our results were nicely reproduced by a theoretical model that incorporates the two-body loss as an imaginary part to the inverse of the scattering volume in the scattering amplitude expression. The observed loss suppression in quasi-two-dimensional traps may provide a promising strategy to realize a $p$-wave superfluid in a system of ultracold atoms.
\end{abstract}

\maketitle

\section{introduction}
Quantum gases with controllable interaction strengths across a $p$-wave Feshbach resonance offer phase transitions between superfluid phases with different symmetries. The transition from a $p_x$ superfluid to a time-reversal breaking $p_x + i p_y$ superfluid as well as the topological quantum-phase transition from a gapped to a gapless $p_x + i p_y$ superfluid state have been theoretically predicted ~\cite{volv, vg, victor, Cheng, fed}. In two dimensions, the vortices in the BCS topological phase of the $p$-wave superfluid with $m_{l}=\pm 1$ trap quasiparticles that obey non-Abelian statistics~\cite{reed} and may have potential applications to realize decoherence-free quantum information processing~\cite{kit}.
In the past, quantum gases of $^{40}$K and $^{6}$Li atoms have been intensively studied across $p$-wave Feshbach resonances to pursue the possibility of achieving a $p$-wave superfluid~\cite{zhang, shunk, gun, gab, inada, fuch, naka, thy, waseem}. However, the observed lifetime of $p$-wave Feshbach molecules so far has been found to be too short to perform conventional evaporative cooling in order to achieve quantum degeneracy~\cite{gab,inada}. The atomic loss near a $p$-wave Feshbach resonance is severe due to three-body recombination or two-body dipolar relaxation, making it challenging to realize $p$-wave superfluidity in a trapped Fermi gas~\cite{reg3,ticknor,chevy,lev}. 

In a two-component Fermi gas, strong two-body dipolar relaxations hinder the control of elastic collisions near a $p$-wave Feshbach resonance. The dipolar losses can be elucidated by including the imaginary part to the inverse scattering volume in the two-body scattering amplitude~\cite{russian20, russian}. The imaginary scattering volume quantifies the measure of the efficiency of dipolar relaxation losses.
Because of enormous inelastic collision losses near a $p$-wave Feshbach resonance, it is important to perform systematic studies to understand the features of the two-body dipolar losses in two-component trapped Fermi gases. 

In this paper, we report the experimental measurement of the two-body dipolar relaxation in a two-component Fermi gas of $^{6}$Li atoms confined in three dimensions (3D) and quasi-2D. We have shown that the three-body contribution in the two-component Fermi gas can be removed by lowering the density. Our experimental results of the two-body relaxation coefficients versus the inter atomic interaction strength and temperature can be systematically explained by including a single parameter for the imaginary part to the inverse of the scattering volume in the scattering amplitude for 3D and quasi-2D atomic Fermi gases.

\section{Two-body relaxation in 3D}
The details of our experimental setup are described elsewhere~\cite{inada,naka}. In brief, we performed evaporative cooling of the atoms prepared in the two lowest atomic states of $\left\vert 1 \right\rangle \equiv \left\vert F=1/2,m_F=1/2 \right\rangle $ and $\left\vert 2 \right\rangle \equiv \left\vert F=1/2,m_F=-1/2 \right\rangle $ in a single-beam optical dipole trap. The two-body relaxation occurs as a spin flip of an atomic state from $\vert 2 \rangle$ to $\vert 1 \rangle$. We kept the temperature of the atoms higher than the Fermi temperature to ensure that the atomic density profile follows the Boltzmann distributions.
We ramped the magnetic field in the vicinity of the $\left\vert 1 \right\rangle-\left\vert 2 \right\rangle$ $p$-wave Feshbach resonance located at $B_0=$~184.9(7)~G. The fluctuation in the magnetic field is suppressed to 8~mG~\cite{naka}. 
The splitting of the two resonances due to the spin-dipole interaction~\cite{ticknor} in $^6$Li is quite small compared with the resonance width. Therefore, we treat the two $p$-wave Feshbach resonances as being completely overlapped with each other.
\begin{figure}[tbp]
\includegraphics[width=3.5 in]{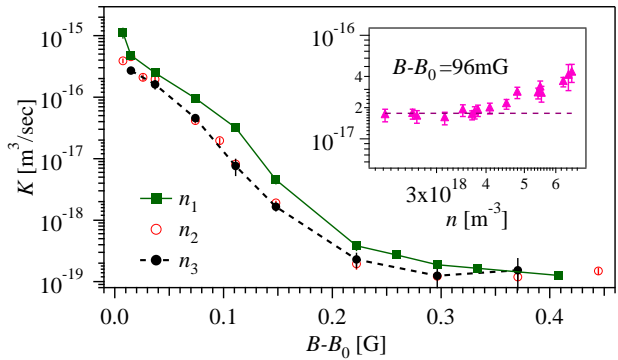}
\caption{The loss rate constant $K$ vs $B-B_0$ at three different atomic densities and temperature of 2.2~$\mu$K. At the higher density $n_{1}$, $K$ deviates from the other two data sets for $n_{2}$ and $n_{3}$, indicating that three-body losses start to contribute at the density $n_{1}$. The inset shows $K$ vs atomic density at $B-B_0 =$~96~mG. The data show a deviation from the two-body loss-dominated regime at $n \approx 4.1\times 10^{18}$~m$^{-3}$.}
\label{fig1}
\end{figure}

To measure atomic losses, we kept the magnetic field $B$ to the desired detuning $B-B_{0}$ and observed the decay of the number of atoms in the trap, which is governed by the following rate equation
\begin{equation}
\frac{\dot{N}}{N}= -\Gamma-\frac{K_2}{2} \left\langle n \right\rangle- K_{3} \left\langle n^2 \right\rangle= -\Gamma-\frac{K}{2} \left\langle n \right\rangle,
\label{rat}
\end{equation}
where $\Gamma$, $K_{2}$, and $K_{3}$ are the one-body, two-body, and three-body loss rate constant, respectively. $\left\langle n \right\rangle$ and $\left\langle n^{2} \right\rangle$ are the mean density and mean square density of both states added up, and $K$ is the total rate coefficient which was determined in the experiment. Since the two-body loss rate is proportional to the density while the three-body loss rate is proportional to the square of the density, the two-body loss dominates the three-body loss at an atomic density lower than some critical value. In this work, we focus on the atomic density dependence of $K$ to limit our measurement to the density region, where the two-body losses dominate the three-body losses. Determining $K_2$ and $K_3$ from the atomic decay is the direct way to show the evolution of dominant loss mechanisms from two-body to three-body losses. However, the shape of the decay curve for two-body loss and three-body loss are not so different and it is quite difficult to identify the loss mechanism from the shape of the decay curve. Instead, we determine the two-body loss coefficients for varying atomic density to see the density dependence of $K_2$ and show the appearance of three-body contribution in the measured loss coefficient as explained later in detail.

Figure~\ref{fig1} shows $K$ versus $B-B_0$ for three different atomic densities ($2^{3/2}\times \left\langle n \right\rangle$), $n_{1} = 6.7\pm0.3\times 10^{18}$~m$^{-3}$ (connected green squares), $n_{2}= 3.7\pm0.5\times 10^{18}$~m$^{-3}$ (red circles), and $n_{3}= 2.5\pm 0.2\times 10^{18}$~m$^{-3}$ (black circles), at a temperature of 2.2~$\mu$K. We varied the atomic density by lowering the initial number of atoms loaded into an optical dipole trap. Although $K$ for $n_{2}$ and $n_{3}$ are consistent with each other, $K$ for $n_{1}$ is higher than $K$ for $n_{2}$ and $n_{3}$.
This is because $n_{1}$ is higher than the critical density above which the three-body losses come into play. The inset of Fig.~\ref{fig1} shows a plot of $K$ versus atomic density at $B-B_0 =$~96~mG. 
It clearly shows a deviation from the two-body loss-dominated regime at the atomic density of $n \approx 4.1\times 10^{18}$~m$^{-3}$, where the three-body loss starts to contribute. 
Below this atomic density, the value of $K$ determined in the experiment can be considered equivalent to $K_2$. 
As a reference, we determined $K_3$ by fitting the atomic decay plots for high atomic density cases with $K_2$ fixed at the value obtained from the low density case, and we get $K_3 \approx 1.5 \pm 0.4 \times 10^{-36}$ ${\rm m}^6/{\rm s}$ at $B-B_{0}=~96 {\rm mG}$.

\begin{figure}[tbp]
\includegraphics[width=3.5 in]{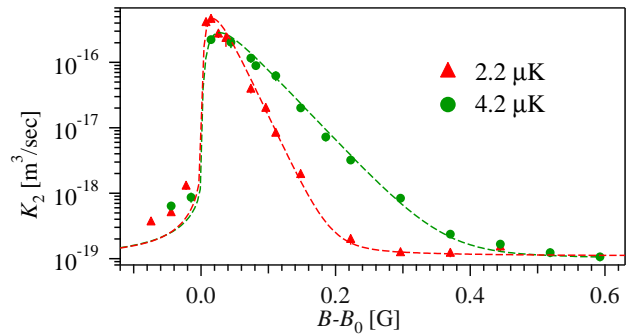}
\caption{The magnetic field dependence of $K_{2}$ at $T=2.2$ (red triangles) and 4.2~$\mu$K (green circles) with density equal to $n_{2}$. The dashed curves represent the theoretical results obtained using Eq.~(\ref{ave1}) by considering the imaginary part of the scattering volume $V_{\rm 1}$ as a fitting parameter.}
\label{fig2}
\end{figure}

Figure~\ref{fig2} shows the measured values of $K_{2}$ with $n \approx n_{2}$ as a function of the magnetic field detuning at $T=2.2$ and 4.2~$\mu$K as indicated by red triangles and green circles, respectively. It can be clearly seen that $K_{2}$ drastically increases at magnetic-field detuning of around zero, and that 
the losses are enhanced over a wider range of magnetic field values for high temperature.
To describe the observed loss feature, we consider the low-energy effective range expansion of the $p$-wave scattering amplitude~\cite{naka20}, which is
\begin{equation}
f(k)=- \frac{k^{2}}{V^{-1}+k_{\rm e} k^2+ik^3},
\label{amp3d}
\end{equation}
where $V=V_{\rm bg}[1-\Delta B/(B-B_{0})]$, with $V_{\rm bg}$ and $\Delta B$ being the background scattering volume and the resonance width, respectively. 
The effective range $k_{\rm e}$ is positive and is assumed to be constant due to its very weak dependence on the magnetic field. Similar to Ref.~\cite{russian}, we included the imaginary part to the inverse of the scattering volume as $1/V=1/V+i/V_{\rm 1}$, where $V_{\rm 1}>0$ and is assumed to be independent of the external magnetic field~\cite{russian}. 
The inelastic loss rate constant can be expressed as $\beta (E) = v_{\rm r} \times \sigma(E)$, where $v_{\rm r}=2\hbar k/m$ is the relative velocity between two atoms of mass $m$ and $\sigma(E)$ is the $p$-wave inelastic scattering cross section for the relative energy $E=m v_{\rm r}^{2}/4$. 
Using the $S$-matrix element notation~\cite{russian}, $\sigma(E)$ can be written as $\sigma(E)=3 \pi (1-\left|S(k)\right|^{2})/k^{2}$ with
\begin{equation}
S(k)=\frac{1/V+k_{\rm e} k^2+i (1/V_1-k^3) }{1/V+k_{\rm e} k^2+i (1/V_1+k^3) }.
\end{equation}
Consequently, an inelastic rate constant (which is twice that of $\beta$) averaged over the Boltzmann distribution can be expressed as follows:
\begin{equation}
K_{2}= \frac{4}{\sqrt{\pi} (k_{\rm B} T)^{3/2}} \int_{0}^{\infty} dE \sqrt{E} e^{-E/k_{\rm B} T} \beta (E).
\label{ave1}
\end{equation}
We adopted the value $V_{\rm bg}\Delta B=-1.8 \times 10^{6} a_{0}^{3}$~[Gm$^{3}$] with $a_{0}$ being the Bohr radius~\cite{luken}. 
In the large scattering volume limit, the effective range can be expressed in terms of scattering parameters from the pole of the scattering amplitude as $k_{\rm e}=-\hbar^{2}/(m V_{\rm bg}\Delta B \delta\mu)$ \cite{victor}, where $\delta\mu=k_{\rm B} \times 111~\mu$K/G is the relative magnetic moment between the molecular state and the atomic state~\cite{fuch} and the effective range is estimated to be $k_{\rm e}\approx 0.14a_{0}^{-1}$.
\begin{figure}[tbp]
\includegraphics[width=3.25 in]{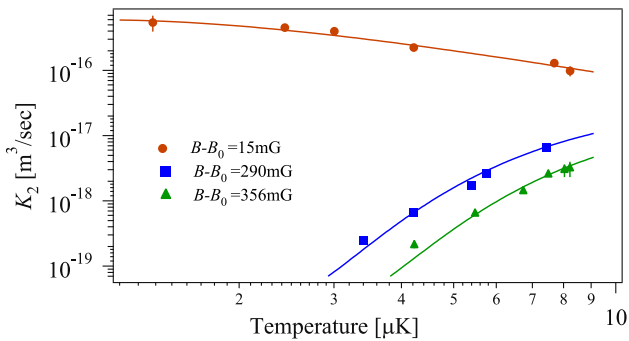}
\caption{The temperature dependence of $K_{2}$ at the different magnetic field detunings. The data clearly show the stark difference in the temperature dependence of $K_{2}$ for the near and far-resonant regimes and is correctly reproduced from Eq.~(\ref{ave1}) as indicated by solid curves.}
\label{fig3}
\end{figure}

The red and green dashed curves in Fig.~\ref{fig2} show the results of data fitting using Eq.~(\ref{ave1}) considering $V_{\rm 1}$ as a free parameter for the data at $T= 2.2$ and 4.2~$\mu$K, respectively \cite{footnote}. Equation~(\ref{ave1}) perfectly reproduces the two sets of experimental data with $V_{\rm 1}=4.85 (0.10)_{\rm stat}(4)_{\rm syst} \times 10^{-21}$~m$^3$. 
The statistical error reflects uncertainty that arises from the fitting error, and the systematic error in fitting is mainly due to uncertainty in the atom number determination.

Next, we fixed the magnetic field detuning and measured the temperature dependence of $K_{2}$. We changed the temperature by changing the final trap depth of the optical dipole trap for evaporative cooling. Figure~\ref{fig3} shows the temperature dependence of $K_{2}$ for different detuning values: $B-B_{0}=$ 15 (circle), 290 (squares) and 356~mG (triangles), while the  density was maintained close to $n_{2}$. The data show a clear difference in the temperature dependence in the near- and far-resonant regimes. 
The solid curves correspond to the theoretical results obtained using Eq.~(\ref{ave1}) with the parameters determined from results shown in Fig.~\ref{fig2}. 
In the near-resonant regime, where $k_T / k_{\rm res}>1$ with $k_T=\sqrt{3m k_{\rm B} T/(2\hbar^2)}$ and $k_{\rm res}=1/\sqrt{k_{\rm e} \left| V \right|}$, $K_{2}$ is approximately expected to depend on $T^{-3/2}$. 
By contrast, in the far-resonant regime, where $k_T / k_{\rm res}<1$, $K_{2}$ is expected to be proportional to $T$~\cite{russian}. The stark difference in the temperature dependence of $K_{2}$ is successfully reproduced in the experiments and theory.

\section{Two-body relaxation in quasi-2D}
We measured the two-body loss coefficient $Q_{2}$ that is defined in a manner similar to that of the 3D case by the rate equation
\begin{equation}
\frac{\dot{N}}{N}= -\Gamma-\frac{Q_{2}}{2} \left\langle n_{\rm 2D} \right\rangle
\label{rat2d}
\end{equation}
in a quasi-2D geometry with $\left\langle n_{\rm 2D} \right\rangle$ being the average 2D atomic density per lattice layer. In this experiment, we experimentally checked that the atomic density is low enough to neglect the three-body loss, and we therefore take into account the two-body loss term in Eq. (\ref{rat2d}). We formed an optical lattice along the quantization axis using a 1064-nm laser whose beam waist was 65~$\mu$m. The quantization axis is parallel to the magnetic field.
In the lattice potential, atoms were distributed in tightly confined isolated 2D layers~\cite{waseem}. 
To completely restrict the motion of the atoms along the lattice direction, we chose an experimental condition such that the axial energy $\hbar \omega_z$, where $\omega_z$ is the confinement frequency along the lattice direction, was higher than the thermal energy $k_{\rm B} T$. Considering both the size of the Brillouin zone and our imaging resolution, we employed the adiabatic band-mapping technique to verify that the population in the excited motional states along the lattice direction is small (less than 8~$\%$)~\cite{miranda}. The atoms in the motional ground state can approach each other only in a side-to-side configuration. Consequently, within each isolated 2D layer, collisions are purely in the $m_{l}=\pm 1$ symmetry~\cite{gun, waseem, tan}.

\begin{figure}[tbp]
\includegraphics[width=3.5 in]{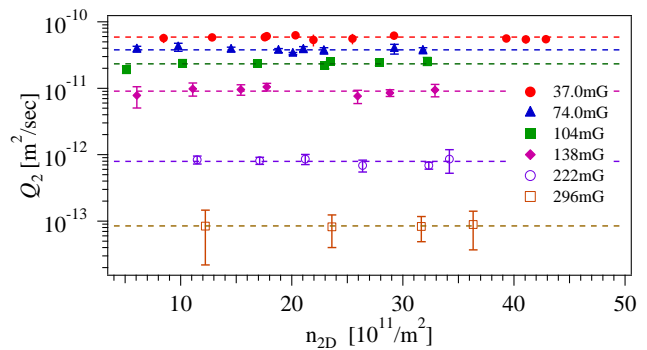}
\caption{The density dependence of $Q_{2}$ for the atoms trapped in quasi-2D at different magnetic field detunings. The dashed lines represent the constant $Q_{2}$, indicating that two-body losses dominate over the three-body losses (see text).}
\label{fig4}
\end{figure}

To measure $Q_{2}$, we monitored the time evolution of the atomic number obtained through absorption images captured along the lattice direction.
An average 2D density per layer is determined by $\left\langle n_{2D} \right\rangle=(N/\alpha) (m \tilde{\omega}^{2}/4 \pi k_{\rm B} T)$~\cite{miranda}, where $N$ is the total number of atoms, $\alpha \approx 133$ is the number of lattice layers, and $\tilde{\omega}=\sqrt{\omega_{x} \omega_{y}}$ is the 2D mean trap frequency.
\begin{figure}[tbp]
\includegraphics[width=3.4 in]{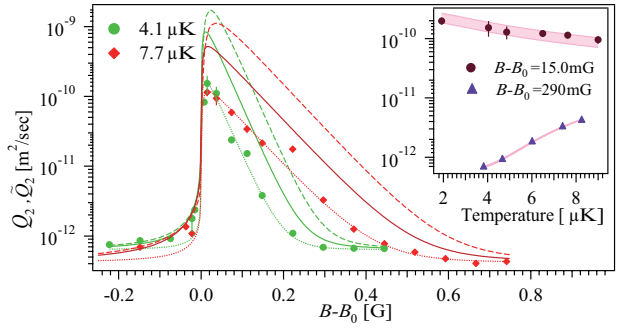}
\caption{Magnetic field dependence of $Q_{2}$ at $T = 4.1$~$\mu$K (green circles) and $T = 7.7$~$\mu$K (red diamonds). The solid and dotted curves represent the results obtained using Eq.~(\ref{ave2}) with an amplitude factor of unity and $\epsilon=0.23\pm0.02$, respectively. The dashed curves show $\tilde{Q_{2}}=K_{2}/(\sqrt{2 \pi} l_{z})$ based on Eq.~(\ref{ave1}), indicating that the loss constant was suppressed in 2D. The inset shows the temperature dependence of $Q_2$. The shaded area represents the results obtained using Eq.~(\ref{ave2}) with $\epsilon =0.23\pm0.02$.}
\label{fig5}
\end{figure}

To ensure that the two-body loss dominates over the three-body loss, we explored the density dependence of $Q_{2}$ at various magnetic field detunings for a fixed temperature of $T = 4.0\mu$K and lattice depth of $V_{\rm lat} = 17$~$E_{\rm rec}$ as shown in Fig.~\ref{fig4}, where $E_{\rm{rec}}=\hbar^2 k^2/(2m) \approx k_{\rm B} \times 1.4$~$\mu$K is the recoil energy of $^{6}$Li. Density was changed similar to the 3D case.
The data clearly indicate that two-body losses dominate in this density region, although an equivalent 3D density $n_{\rm 2D}/(\sqrt{\pi} l_{z})$ is approximately one order higher than densities of Fig.~\ref{fig1}. Here, $l_{z}=\sqrt{\hbar/m\omega_{z}}$ is the harmonic oscillator length with $\omega_{z} = 2 \pi \times 245$~kHz. 
Since the three-body loss does not appear at $n_{\rm 2D} \approx 3.5 \times 10^{12} {\rm m}^{-2}$ for $B-B_0 \approx 100 {\rm mG}$, the upper limit of $K_3$ is estimated to be $2 \times 10^{-37} {\rm m}^6/{\rm s}$. This value is one order smaller than the 3D case ($1.5 \times 10^{-36} {\rm m}^6/{\rm s}$ as discussed already). In this sense the three-body losses are observed to be less important in 2D and the higher suppression of the three-body loss in 2D is also consistent with the prediction by Ref.~\cite{fed}. Since three-body losses are less important in 2D, to overcome the issue of the three-body loss in a high density regime for a two-component Fermi gas, it might be useful to confine the atoms in the quasi-2D.
In Fig.~\ref{fig5}, the green circles show $Q_{2}$ for $n_{\rm 2D} = 3.3 \times 10^{12}$~m$^{-2}$, $T = 4.1 \mu$K, and $V_{\rm lat} = 22$~$E_{\rm rec}$ ($\omega_{z} = 2 \pi \times 275$~kHz) while 
red diamonds show $Q_{2}$ for $n_{\rm 2D} = 3.8\times 10^{12}$~m$^{-2}$, $T = 7.7~\mu$K, and $V_{\rm lat}=53$~$E_{\rm rec}$ ($\omega_{z} = 2 \pi \times 428$~kHz).

Similar to the 3D case, we explain the two-body loss considering the quasi-2D $p$-wave scattering amplitude for two particles with a relative momentum $q$~\cite{prico, cai},
\begin{equation}
f(q)= \frac{4 q^{2}}{\frac{1}{A}+q_{\rm e} q^2-(2 q^{2}/\pi)ln(l_{z} q)+iq^2},
\label{amp2d}
\end{equation}
where, $A=(3 \sqrt{2 \pi} l_{z}^{2}/4)/(l_{z}^{3}/V+k_{\rm e} l_{z}/2-0.06553)$ and $q_{\rm e}=(4/3 \sqrt{2 \pi})[l_{z} k_{\rm e}-1.13015]$ are the 2D scattering area and the effective range, respectively.

The inelastic cross section for quasi-2D can be written in terms of the $S$ matrix as $\sigma_{\rm in}^{\rm 2D}=(2/q)[1-\left|S_{\rm 2D}(q)\right|^{2}]$, where the 2D scattering amplitude is related to the $S$ matrix as $f(q)=2i[S_{\rm 2D}(q)-1]$. 
Similar to the analysis of the 3D case, we included the imaginary part to the inverse of the scattering volume $V_{\rm I}$ which yields an imaginary scattering area: $A_{\rm I}=3 \sqrt{2 \pi} V_{\rm I}/4 l_{z}>0 $. 
Then, by taking the thermal average, the 2D inelastic loss rate constant, which is twice the value of $\beta_{2D}=v_{\rm r} \times \sigma_{\rm in}^{2D}$, can be expressed as follows:
\begin{equation}
Q_{2}= \frac{1}{(k_{\rm B} T)} \int_{0}^{\infty} dE \beta_{2D} \times \exp (-E/k_{\rm B} T).
\label{ave2}
\end{equation}
In Fig.~5, the solid curves represent the result obtained using Eq.~(\ref{ave2}) with the parameters determined from the 3D measurement. 
Compared to the case of the 3D trap, the theoretical curves do not match with the experimental results and differ by factor of four. However, by taking the amplitude factor of $\epsilon=0.23\pm0.02$, the curves match nicely with experimental results for $T = 4.1 \mu$K and $T = 7.7 \mu$K as shown by dotted curves. 
Even if we take into account the systematic uncertainty (60 $\%$ or factor of 2.5) associated with atomic numbers, trap frequencies, temperatures, and lattice depth $V_{\rm lat}$, the deviation of experimental data from theory curves can not be explained. 
The reason for a factor $\epsilon$ smaller than unity comes from the dipolar loss suppression in the $m_{l}=-1$ collision channel. Since the atoms are confined in a quasi-2D trap perpendicular to the quantization axis, atoms can only collide in the $m_{l}=\pm 1$ configurations~\cite{gun, waseem, tan}. 
The $m_{l}=-1$ channel has a four orders of magnitude smaller dipolar loss coefficient than the $m_{l}=+1$ channel because of the angular momentum conservation~\cite{chevy}. 
Therefore, the number of channels that contribute to the dipolar loss is reduced by a factor of two in the lattice potential ($m_{l}=+1$ channel contributes in 2D case and $m_{l}=0$ and $m_{l}=+1$ channels contribute in 3D case). The overestimation of the number of collision channels may cause $Q_{2}$ to be overestimated by a factor of two, and this may be a part of the reason of the uncertainty. 
However, this scenario needs to be tested by performing the same measurement with an optical lattice which is aligned perpendicular to the quantization axis to make both the $m_{l}=0$ and $m_{l}=\pm 1$ channels active.
In the non-degenerated regime, the atomic loss rate constant determined in the 3D measurement can be scaled to the atomic loss rate constant in 2D by $\tilde{Q_{2}}=K_{2}/(\sqrt{2 \pi} l_{z})$~\cite{miranda} unless additional properties of the loss mechanism are modified by the lattice confinement. We compared $\tilde{Q_{2}}$ with values of $Q_2$ using Eq.~(\ref{ave1}) as shown by dashed curves in Fig~\ref{fig5} which clearly shows that the loss rate constant is suppressed in quasi-2D compare with the 3D case. This suppression is predicted theoretically in Ref.~\cite{russian}. Our experimental results (markers in Fig.~5) show even smaller dipolar loss coefficients than those which were theoretically predicted (solid curves in Fig.~5).

Furthermore, the temperature dependence of $Q_{2}$ was measured at small and rather large magnetic field detuning conditions, as shown in the inset of Fig.~\ref{fig5}.
Since the two-body loss rate constant is insensitive to the confinement strength, we scanned the temperature of the quasi-2D Fermi gas by changing the confinement strength $\omega_z$ from $2\pi \times 75$~kHz to $2\pi \times 500$~kHz in the range shown in the inset of Fig.~\ref{fig5}.
The width of the shaded area indicates the variation in the values of $Q_2$ by changing $\omega_z$ from $2\pi \times 75$ to $2\pi \times 500$~kHz. 
A clear difference in the temperature dependence of $Q_{2}$ can be observed between the near- and far-resonant conditions. 
In the near-resonant regime, $q_{\rm T}/q_{\rm res}>1$ with $q_{\rm T}=\sqrt{m k_{\rm B} T/(\hbar^2)}$ and $q_{\rm res}=1/\sqrt{q_{\rm e} \left| A \right|}$, the loss rate constant approximately behaves as $T^{-1}$ compared with the 3D case in which the loss rate constant has a $T^{-3/2}$ dependence.

\section{conclusion and outlook}
In conclusion, we experimentally performed systematic studies of the two-body loss of atoms in the vicinity of a $p$-wave Feshbach resonance using a two-component Fermi gas of $^6$Li atoms confined in 3D and quasi-2D traps. The observed two-body loss rate constants as functions of temperature and magnetic field can be explained accurately by introducing an imaginary part to the inverse of the scattering volume. Our findings support the prediction of the loss suppression in 2D and we experimentally observed an even higher suppression than that predicted theoretically. Since further suppression of the dipolar loss in one dimension has been discussed theoretically~\cite{russian}, confining the atoms in one dimension may provide the way to achieve $p$-wave superfluid in cold atom systems. The elastic collisional properties of the atoms in the low dimension as well as the inelastic collisional properties studied in the current work are also important toward realization of $p$-wave superfluid.

\section*{ACKNOWLEDGMENT}
We acknowledge support from a Grant-in-Aid for Scientific Research on Innovative Areas (Grant No. 24105006). M.W. acknowledges the support received from the Japanese Government Scholarship (MEXT).


\begin{thebibliography}{90}
\bibitem{footnote} The two-body relaxation is slow at the large detuning region (flat region), we add the one-body contribution $\Gamma /n$ to Eq. ~(\ref{ave1}) to fit the data.
\end{thebibliography}
\end{document}